\documentclass[aps,prl,twocolumn,superscriptaddress,amsmath,floatfix]{revtex4-2}

\usepackage{graphicx}% Include figure files
\usepackage{dcolumn}% Align table columns on decimal point
\usepackage{bm}% bold math
\usepackage{amssymb}
\usepackage{hyperref}
\usepackage{multirow}
\usepackage{color}
\usepackage{braket}

\begin{document}

\title{Phase Separation in a system of Brownian inertial rotors}

\author{Lucio Mauro Carenza}
%\orcidlink{0009-0001-9128-8200}
\affiliation{Dipartimento Interateneo di Fisica, Universit\`a degli Studi di Bari, via Amendola 173, Bari, I-70126, Italy}
\affiliation{INFN, Sezione di Bari, via Amendola 173, Bari, I-70126, Italy}

\author{Pasquale Digregorio}
%\orcidlink{0000-0001-6567-9382}
\affiliation{Dipartimento Interateneo di Fisica, Universit\`a degli Studi di Bari, via Amendola 173, Bari, I-70126, Italy}
\affiliation{INFN, Sezione di Bari, via Amendola 173, Bari, I-70126, Italy}

\author{Massimiliano Semeraro}
%\orcidlink{0000-0001-8273-4232}
\affiliation{Laboratoire de Physique Théorique et Modélisation, CNRS UMR 8089, CY Cergy Paris Université, F-95032 Cergy-Pontoise Cedex, France}

\author{Antonio Suma}
%\orcidlink{0000-0002-5049-9255}
\affiliation{Dipartimento Interateneo di Fisica, Universit\`a degli Studi di Bari, via Amendola 173, Bari, I-70126, Italy}
\affiliation{INFN, Sezione di Bari, via Amendola 173, Bari, I-70126, Italy}

\author{Ignacio Pagonabarraga}
%\orcidlink{0000-0002-6187-5025}
\affiliation{Universitat de Barcelona Institute of Complex Systems (UBICS), Universitat de Barcelona, 08028 Barcelona, Spain}
\affiliation{Departament de F\'{\i}sica de la Mat\`eria Condensada, Universitat de Barcelona, Carrer de Mart\'{\i} i Franqu\'es 1, 08028 Barcelona, Spain}

\author{Giuseppe Gonnella}
%\orcidlink{0000-0002-1829-4743}
\affiliation{Dipartimento Interateneo di Fisica, Universit\`a degli Studi di Bari, via Amendola 173, Bari, I-70126, Italy}
\affiliation{INFN, Sezione di Bari, via Amendola 173, Bari, I-70126, Italy}

\date{\today}

\begin{abstract}
We uncover an emergent phase separation in a system of inertial dumbbells put into persistent rotation by a constant applied torque, breaking chiral symmetry, while interacting with each other through repulsive hard-core interactions.
We numerically determine the phase diagram of the model, showing that there exists a binodal and a spinodal region where the system separates into a dense liquid and an ordered phase, which we identify as hexatic.
An analysis of the system pressure unveils that the mechanism underlying the phase separation is heavily controlled by inertia.
Locally injected kinetic energy is steadily stored into the system in a way that depends on density and that, at large inertia, results in a hot dilute and cool dense state, leading to a generically unstable homogeneous phase.
The interplay between dumbbells' arrangement into clusters and chiral forces produces a spontaneous rigid rotation of the clusters around their centers of mass.
\end{abstract}

\maketitle

%{\it Introduction}.
Phase transitions and self-organization are reasonably understood in thermal systems, where a free-energy description is available. 
Moving to non-equilibrium systems, a surprisingly rich variety of new phenomena and phase behavior is emerging, sometimes even without a solid theoretical description. 
Important examples, such as flocking~\cite{Viseck_1995,chate_flocking} and motility-induced phase separation (MIPS)~\cite{tailleur2008,fily2012,digregorio2018}, occur in active systems, where constituents are provided with a constant energy input, transformed into directed translational motion.
In MIPS, thermal self-propelled particles display phase separation into a gaseous and a dense phase even in the absence of attractive interactions. 

Another class of active systems are those where local driving consists of torques acting on each individual unit. 
Ensembles of biological active rotors are known to organize in ordered structures~\cite{lenz2003,riedel2005}.
Striking examples are  stable colonies of spinning {\it Volvox algae}~\cite{drescher2009} or 2D crystals of rotating cells of { \it Thiovulum majus} bacteria~\cite{petroff2015}. 
Self-rotating particles can also be engineered in experiments where the rotations are controlled by external driving~\cite{kokot2017,modin2023,fran_spinners_commphys,fran_spinners_PRR,fran_spinners_pof}, spanning a wide range in mass and size, from micrometer-size colloids to centimeter-size massive and inertial particles. 
Several of these studies, also supported by the analysis of theoretical models, focused on the combined effect of chirality and hydrodynamics on ordering~\cite{snezhko2016,goto2015,yuan2024,soni2019,wang2019,shen2020}. 
Differently, in dry systems~\cite{tsai_chira_gran,workamp_airspinners,yang_exp_PRE,han_fluctuating_2021} it is the friction between rotating units that produces non-trivial collective behaviors. 
A chirality-breaking phase separation in mixtures of rotors has been observed in~\cite{nguyen2014,scholz_spinning_robots,lopez2022}. 
Transitions to  non-homogeneous states, even induced by rotational activity, have also  been identified~\cite{joshi2022,caprini_bubble,digregorio2025_spinners}. 
In~\cite{joshi2022}, rotating magnetic fields are shown to drive steady-state vapor-liquid coexistence in a suspension of superparamagnetic colloids.
In~\cite{caprini_bubble, digregorio2025_spinners}, the formation of patterns with spatially ordered bubbles that 
coexist with the liquid is attributed to a complex interplay between chiral interactions and particle inertia. 
This segregation is interrupted with a typical finite size of stationary bubbles.

\begin{figure}[t!]
    \centering 
    \includegraphics[width=\columnwidth]{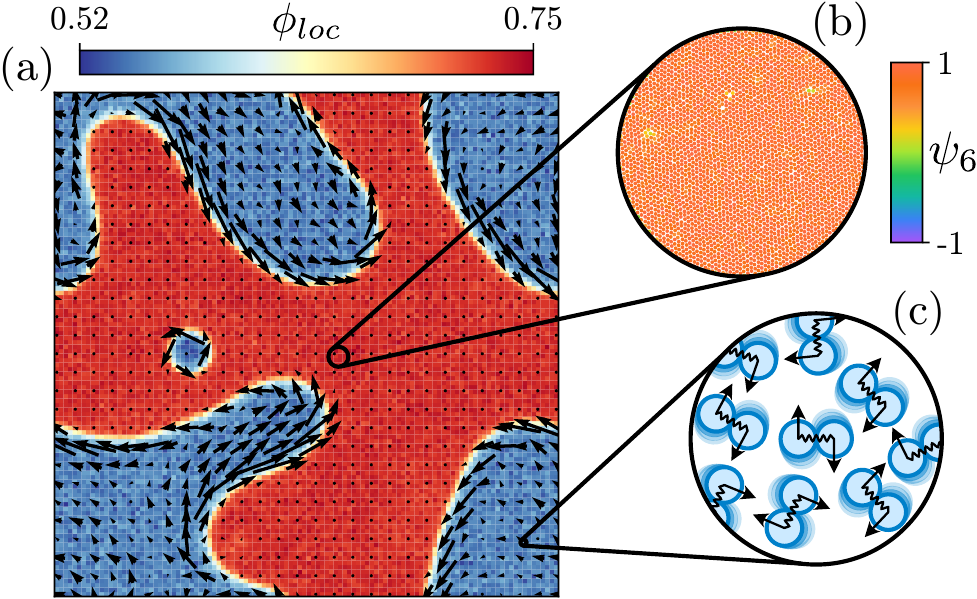}
    \caption{
    (a) A configuration  of $N_d=2 \cdot 1024^2$ dumbbells, evolving according to Eq.~\eqref{eq:langevin}, during coarsening from the homogeneous state into the separated state, for $Pr=5 \cdot{10^{-3}}$, $Pe=5$ and global surface fraction $\phi=0.65$.
    Colors represent local surface fraction $\phi_{loc}$, while the black arrows represent local velocities.
    (b) An enlargement of the dense phase, showing the projection $\psi_6$ of the local hexatic order parameter $\psi_{6,i}$ [see definition in the text] on the global mean orientation.
    (c) An enlargement of the disordered phase, showing internal bonds and active torques acting on the dumbbells.
    } 
    \label{fig:model}
\end{figure}

In this Letter, we show that the combined effects of chirality and inertia lead to a distinct  kind of macroscopic phase transition.
We consider a system of spinners consisting of thermal dumbbells, rotating on a substrate, as sketched in Fig.~\ref{fig:model}(c). 
A similar model was studied in~\cite{ni_PNAS19}, to show the emergence of hyperuniform states~\cite{lei2025,ni_PNAS19} and in~\cite{poggioli_dumbbells_PRL23,poggioli_limmer_chiral} in relation to rheological behavior. 
Here, we show that,  at sufficiently high density, intrinsic local rotations induce hexatic-liquid melting, characterized by a binodal region within a density range that increases with the intensity of rotation.
An example of  configuration with coexisting phases is shown in Fig.~\ref{fig:model}(a).
Inertia proved to be essential for the emergence of this phenomenon.
We also investigate the kinetics of domain coarsening, characterized by a diffusive growth's law for the typical domain size $R(t)=t^{1/3}$. 
This phase separation is accompanied by the emergence of rigid cluster rotations and edge currents, as  shown in Fig.~\ref{fig:model}(a).

%{\it Model}.
We consider a 2D system of $N_d$ dumbbells---diatomic molecules composed of two identical circular beads of mass $m$ and diameter $\sigma$---moving according to the following Langevin equations:
\begin{equation}
\label{eq:langevin}
    m \ddot{\mathbf{r}}_i = -\gamma \dot{\mathbf{r}}_i - \boldsymbol{\nabla}_i U + \mathbf{F}_{i,a} + \sqrt{2 \gamma k_BT} \ \boldsymbol{\xi}_i \ \rm{,}
\end{equation}
with $\mathbf{r}_i$ the position of the $i$-th bead.
Here, $i=1, \dots, N$, with $N=2N_d$.
All beads are coupled with a stochastic thermal bath of temperature $T$, with friction coefficient $\gamma$ and $\boldsymbol{\xi}$  uncorrelated Gaussian noise with zero mean and unit variance.
The interparticle potential, $U = U_{\rm{WCA}} + U_{\rm{FENE}}$, is the sum of a purely repulsive Weeks-Chandler-Anderson (WCA) potential~\cite{Weeks_1971}, 
$U_{\rm{WCA}}(r) = 4\varepsilon \bigl[ (\sigma/r)^{12} + (\sigma/r)^6 \bigr]$, 
cut at its minimum $r_c = 2^{1/6}\sigma$, acting between every pair of beads, and a FENE potential~\cite{fene1990},  
$U_{\rm{FENE}}(r) = K R_0^2 \ln \bigl[ 1-(r/R_0)^2 \bigr]$, between the beads of the same dumbbell, which enforce the molecular bond.
The force $\mathbf{F}_{i,a}$ has fixed magnitude $F_a$ for all beads and is perpendicular to the dumbbell longitudinal axis, with $\mathbf{F}_{i,a}=-\mathbf{F}_{j,a}$ within the same dumbbell [see Fig.~\ref{fig:model}(c)].
In units of $m$, $\sigma$ and $\varepsilon$, we set the bath temperature $T=10^{-2}$, while varying $\gamma$ to tune the particles' inertia.
For the FENE potential, we use $K=30$ and $R_0=1.5$.
We use LAMMPS~\cite{LAMMPS} to perform Molecular Dynamics (MD) simulations in a square box of side $L$, evolving particles according to Eq.~\eqref{eq:langevin}.
We change $L$ to adjust the global surface fraction of the system $\phi=N\pi\sigma^2/4L^2$.

Under the action of the active torque $\tau_a=\sigma F_a$,
here chosen clockwise, a free dumbbell undergoes translational diffusion ($D=k_BT/2\gamma$) and rotational diffusion ($D_r=2k_BT/\gamma \sigma^2$) with a drift angular velocity $\omega_a=2F_a/\sigma\gamma$ [see Appendix A].
Using $D_r^{-1}$ as timescale, we derive a dimensionless version of Eq.~\eqref{eq:langevin} in terms of a set of dimensionless parameters [Appendix A]. 
We define the  P\'eclet number $Pe=\sigma F_a/2k_BT$, as in similar active systems, e.g. in~\cite{poggioli_dumbbells_PRL23}, measuring the ratio between advective rotational transport and diffusion. 
An active Reynolds number can also be defined as $Re=mF_a/\sigma \gamma^2$, representing the ratio between inertial and rotational characteristic timescales $\tau_I/\omega_a^{-1}$.
Then, we define a Prandtl number $Pr=Pe/Re=\gamma/m D_r$, formally analogous to the one defined for convective thermal systems, measuring the ratio between the translational momentum damping rate and the rotational diffusion rate.
While $Pe$ controls rotational activity, $Pr$ is proportional to the translational inertia timescale $\tau_I=m/\gamma$, so that the overdamped limit of Eq.~\eqref{eq:langevin} is recovered in the limit $Pr=\infty$.

\begin{figure}[t!]
    \centering 
    \includegraphics[width=\columnwidth]{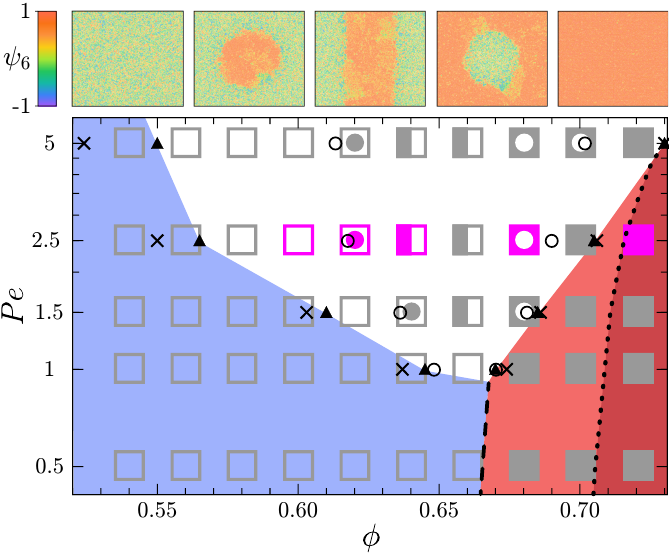}
    \caption{Phase diagram of the model, in the regime of large inertia at $Pr=200$.
    The blue and red regions represent the homogeneous liquid and the ordered phases, respectively.
    Black dashed and dotted lines indicate the liquid-hexatic and hexatic-solid transition, respectively.
    The phase separated region (white area) is delimited by the binodal, indicated with black triangles representing the measured coexisting densities.
    Black crosses and circles indicate the binodal and spinodal densities, respectively, as extracted by the Maxwell construction and the location of the maximum and minimum of the measured pressure.
    Accordingly, symbols indicate the typical state morphology: \raisebox{-1pt}{\includegraphics[height=0.8\baselineskip]{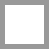}} represents a homogeneous liquid, \raisebox{-1pt}{\includegraphics[height=0.8\baselineskip]{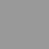}} an ordered state, \raisebox{-1pt}{\includegraphics[height=0.8\baselineskip]{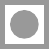}} a dense droplet immersed in the liquid, \raisebox{-1pt}{\includegraphics[height=0.8\baselineskip]{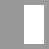}} a dense stripe, \raisebox{-1pt}{\includegraphics[height=0.8\baselineskip]{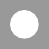}} a liquid droplet in the dense phase. 
    Simulations configurations in these states, corresponding to the symbols highlighted in violet, are shown on the top. 
    }
    \label{fig:phase_diagram}
\end{figure}

In the following, we study the phase diagram at fixed $\gamma=10^{-2}$, hence $Pr=5 \cdot 10^{-3}$, varying $F_a$ to change $Pe$.
We explore a range of P\'eclet such that $Re \geqslant 10^2$, thus translational and rotational motion are highly inertial.
In the absence of active torque, this system undergoes a continuous liquid-hexatic and a hexatic-solid phase transition for increasing global density, corresponding to the emergence of quasi-long range hexatic and translational order, respectively~\cite{cugliandolo_dumbbells_2017,petrelli2018,krauth_continuous_15}.
We observe that this behavior is retained for small active torques, up to $Pe \simeq 1$, as shown in the phase diagram of Fig.~\ref{fig:phase_diagram}, while the critical density for both transitions increases slightly with  increasing $Pe$.

As the spinning velocity increases above $Pe \gtrsim 1$, we observe the emergence of liquid-hexatic phase separation, identified with the white region in Fig.~\ref{fig:phase_diagram}.
The separation between coexisting densities spreads as $Pe$ increases, as shown by the binodal curve, identified with black triangles in Fig.~\ref{fig:phase_diagram}.
These are measured from the location of the two peaks in the bimodal probability distributions of the local density [see Appendix B].
Inertia plays a major role in the phase behavior of the system.
Indeed, increasing $Pr$ (by increasing $\gamma$ in Eq.~\eqref{eq:langevin}) progressively suppresses phase separation and eventually drives the system into a homogeneous state at all $\phi$ and $Pe$.
This is explicitly shown in the $Pr-Pe$ phase diagram in Appendix C, exploring a wide range of Prandtl numbers, which includes a small Reynolds regime, where phase separation is not observed. 

Compatibly with the ordering scenario at small $Pe$, the dense phase at high $Pe$ shows hexatic order.
In Fig.~\ref{fig:phase_diagram}, this is evidenced by the behavior of the projection on the horizontal axis of the local hexatic order parameter $\psi_{6,i}=n_i^{-1}\sum_{j=1}^{n_i} e^{{\rm{i}}6\theta_{ij}}$, where the sum runs over the $n_i$ first neighbors of particle $i$ and $\theta_{ij}$ is the relative bond angle referred to the $x$-axis.
The high-density region of the phase diagram also includes the hexatic-solid transition line [dotted line in Fig.~\ref{fig:phase_diagram}] emerging from the $Pe=0$ axis. 
This has been calculated from the behavior of the hexatic and translational correlation functions, that is detailed in the Supplemental Material (SM)~\cite{sm}.

The binodal region of Fig.~\ref{fig:phase_diagram} shows  a rich phase behavior, which we first explore with simulations starting from a homogeneous disordered state.
Within the binodal, to the right of the liquid branch, there is a range of densities where the system remains in the homogeneous phase.
Moving to larger densities the homogeneous phase becomes  unstable, favoring a sudden decomposition of the system into two phases, which then coarsen to reach a complete phase separated state.
This behavior is compatible with the presence of a spinodal region, outside which the homogeneous state is metastable.
Depending on the abundance of one phase with  respect to the other, the stationary state is indicated by representative symbols described in Fig.~\ref{fig:phase_diagram}.

\begin{figure}[t!]
    \centering 
    \includegraphics[width=\columnwidth]{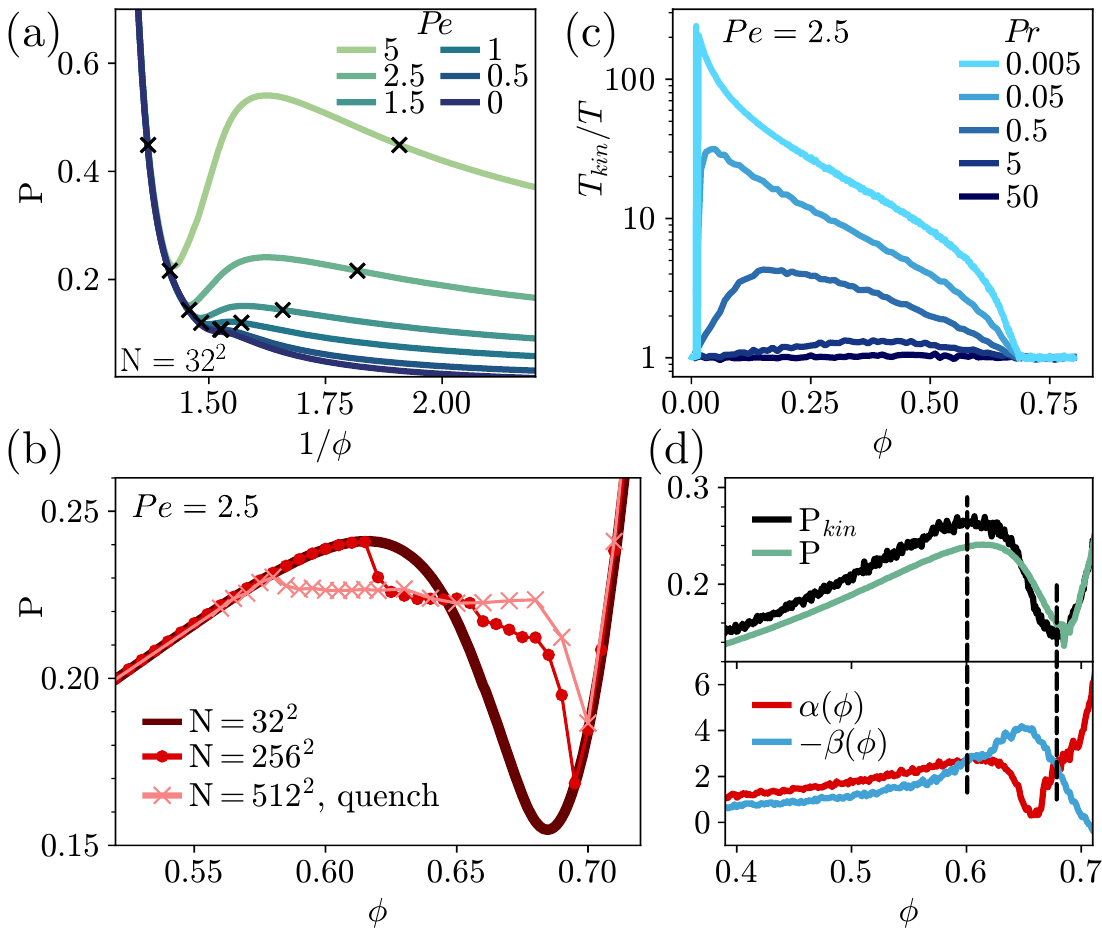}
    \caption{
    (a) IK pressure for a system of $N=1024$ beads and different P\'eclet numbers, showing the Maxwell equal area construction for $Pe\geq1$.
    (b) Pressure as a function of $\phi$ for a system of $N=1024$ beads (green line), $N=256^2$ (light blue dots), $N=512^2$ quenched from the separated state (blue crosses).
    (c) Kinetic temperature as a function of $\phi$ for a system of 1024 beads at $Pe=2.5$ and different values of $Pr$.
    (d) (Top) Comparison between the IK and kinetic pressure for a system of 1024 beads. (Bottom) The two contribution $\alpha(\phi)$ and $\beta(\phi)$ to the pressure slope.
    }
    \label{fig:pressure}
\end{figure}

In order to further characterize phase equilibria within the binodal region, we measure the Irving-Kirkwood (IK) pressure~\cite{ik_pressure} $P=\frac{1}{V} \sum_{i=1}^N \bigl[ m v_i^2+\bm{r}_i \cdot \bm{F}_i \bigr]$, where $v_i$ is the instantaneous particle velocity and $\bm{F}_i$ is the total force on particle $i$.
It includes contributions from the WCA potential and the molecular bond.
The active torque has a null contribution to IK formula---{for two particles $i,j$ belonging to the same dumbbell $(\bm{r}_{i}-\bm{r}_{j}) \cdot \bm{F}_{i,a}=0$---though it contributes to the pressure indirectly through the particles' coordinates, which are modified by the presence of the torque.

The pressure $P$, averaged over several independent configurations, is reported in Fig.~\ref{fig:pressure}(a) for a small system of $N=1024$ beads at different $Pe$s.
Being particle segregation prevented for this small system size, this represents the bare pressure of a homogeneous state, without interfaces, referred to as \textit{homogeneous pressure} $P_{h}$, which  will be used for comparison with the pressure in the phase separated state.
For $Pe \geq 1$, $P_{h}$ exhibits a Van der Waals loop, with a region of negative compressibility at intermediate densities.
This region is limited by  circles in Fig.~\ref{fig:phase_diagram}, and corresponds to the extension of the spinodal previously discussed. 
The Maxwell construction computed in these cases returns values of the binodal densities compatible with the ones measured in the corresponding separated states [black crosses reported in Fig.~\ref{fig:phase_diagram}].
Small deviations could be ascribed to finite-size effects.

In a bigger system of $N=256^2$ beads (teal circles in Fig~\ref{fig:pressure}(b)), $P$ coincides with the homogeneous pressure outside the region of negative compressibility, where the system is in homogeneous states (symbols \raisebox{-1pt}{\includegraphics[height=0.8\baselineskip]{figures/empty.pdf}} and \raisebox{-1pt}{\includegraphics[height=0.8\baselineskip]{figures/filled.pdf}} in Fig.~\ref{fig:phase_diagram}). 
This behavior, with a sudden drop after the maximum of $P_h$, again suggests the presence of a spinodal region where the homogeneous state is unstable and the system spontaneously separates.
To confirm the metastability of the homogeneous state between the spinodal and the binodal, we performed additional simulations with a quench from the separated state to lower densities, across the spinodal, for a system of $N=512^2$ (blue crosses in Fig.~\ref{fig:pressure}(b)), showing that the system remains in the separated state.

\begin{figure}[t!]
    \centering 
    \includegraphics[width=\columnwidth]{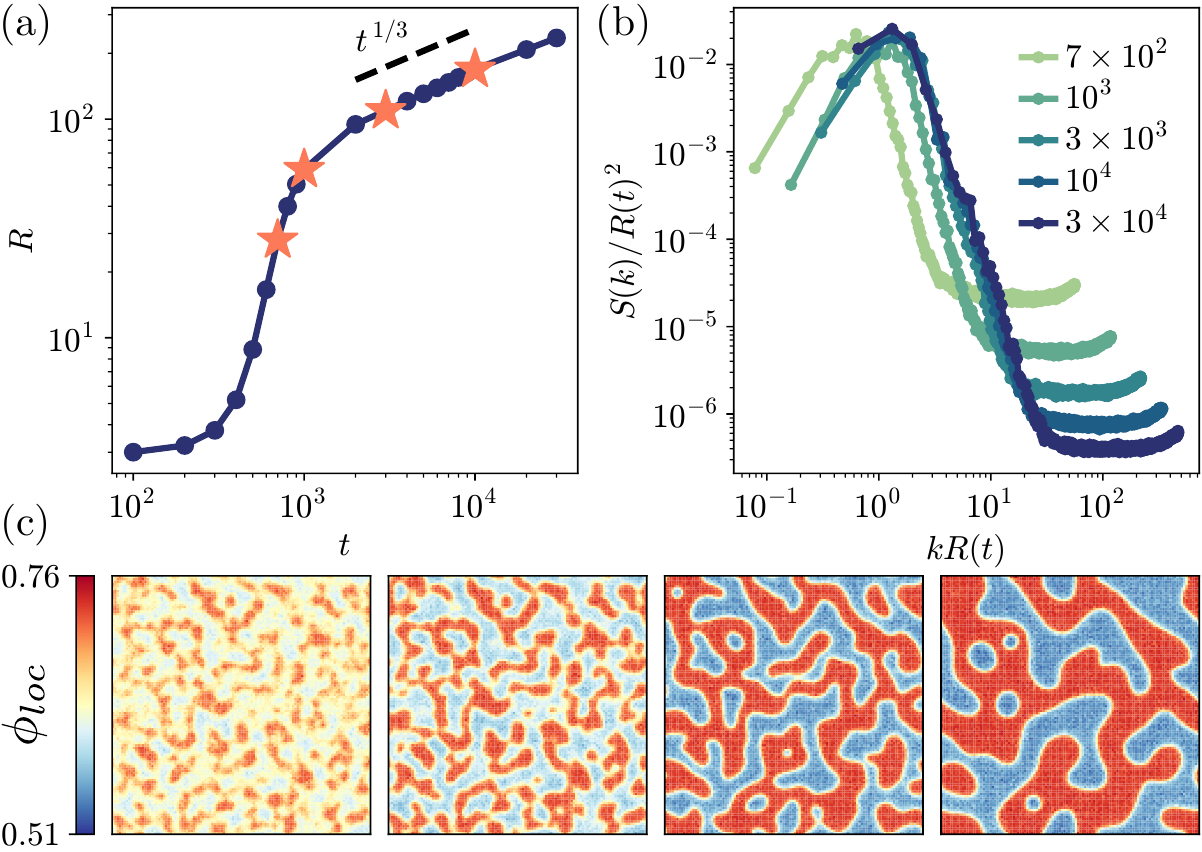}
    \caption{
    (a) Growth of the domain characteristic length $R(t)$  after a quench from the disordered state, averaged over 3 independent simulations.
    (b) Rescaled structure factor at different times.
    (c) Snapshots of the system, representing the local surface fraction, at  the times indicated by red stars in panel (a).}
    \label{fig:ps_kinetic}
\end{figure}

In the small-$Pr$ regime, the dumbbells possess translational inertia, which allows the system to steadily store kinetic energy.
This is captured by the translational kinetic energy $T_{kin}=\sum_{j=1}^{N_d} v_{CM,j}^2/2$, shown in Fig.~\ref{fig:pressure}(c).
At $Pr=5 \cdot 10^{-3}$, while $T_{kin} \sim T$ for $\phi \sim 0$, it rapidly increases at small density, as rotational kinetic energy - injected in the system by the active torque - is transformed into translational kinetic energy through collisions.
This amount of energy is retained by the dumbbells due to finite inertia before it is dissipated over an average time $\tau_I$.
At high density, on the other hand, the average free flight time is much shorter than $\tau_I$, preventing dumbbells from gaining rotational kinetic energy  between collisions and $T_{kin} \sim T$. 
Overall, collision rate and inertial timescale compete in a way that depends on the system density, with $T_{kin}$ showing the non trivial behavior of Fig.~\ref{fig:pressure}(c).
This non-monotonous net energy gain is progressively suppressed as $Pr$ increases and disappears if inertia is negligible, because the kinetic energy gained at collisions is readily dissipated and $T_{kin}=T$, regardless of the system density.

The density-dependent kinetic temperature can be used for a different evaluation of the pressure.  
Generalizing the equilibrium state equation, one can define  $P_{kin}=\phi T_{kin} \chi$.
The quantity $\chi$ takes into account deviations from the ideal gas due to particle interactions, thus it is a strictly increasing function of $\phi$ and $\chi \to 1$ in the ideal gas limit.
Remarkably, measuring $\chi$ in the equilibrium system and computing $P_{kin}$ at $Pe=2.5$ shows a quantitative agreement with the Van der Waals loop of pressure [Fig.~\ref{fig:pressure}(d)]. 
The slope of $P_{kin}$ consists of two contributions: $\alpha(\phi)=T_{kin}\frac{{\rm{d}}(\phi\chi)}{{\rm{d}}\phi}$, controlled by steric repulsion, and $\beta(\phi)=\phi\chi\frac{{\rm{d}}T_{kin}}{{\rm{d}}\phi}$ which only appears with non zero inertia.
While $\alpha$ is always positive, $\beta$ takes negative values at intermediate densities.
For small enough $Pr$ the two sum up to a decreasing pressure over a finite density range, as shown in Fig.~\ref{fig:pressure}(d) for the case $Pr=5 \cdot 10^{-3}$, $Pe=2.5$. 
This clarifies the crucial role of inertia in the emergence of the above separation phenomenon.
It modulates non-monotonously the local energy injection through the active torque, making the homogeneous phase mechanically unstable, in favor of a phase separated state.

A similar mechanism for a non-monotonic pressure has been identified for inertial rotating disks interacting with transverse contact forces~\cite{digregorio2025_spinners}.
However, in~\cite{digregorio2025_spinners}, as well as in other works on chiral disks~\cite{caprini_bubble,lintuvuori_cavitation}, phase separation is arrested, with gaseous bubbles of finite size coexisting with a dense phase.
Here, we observe a genuine phase separation as it also emerges from the analysis of domain coarsening in a system with $N=1024^2$, after a quench from a disordered state into the spinodal region at $Pe=2.5$ and $\phi=0.65$.
Fig.~\ref{fig:ps_kinetic}(a) shows a typical spinodal phase separation process, with the final stage characterized by a growth law $R(t) \sim t^{1/3}$ for the average  domain size, evaluated as the inverse first moment of the structure factor $S(\bm{k},t)$.
This is the universal dynamical exponent of phase separation kinetics with underlying diffusive dynamics~\cite{bray_ordering}.
Interestingly, this dynamical exponent differs from the one measured for MIPS in self-propelled dumbbell systems~\cite{caporusso_dumbbells2d,semeraro_entropy25}, indicating that this phase separation is significantly different from MIPS.
Moreover, in Fig.~\ref{fig:ps_kinetic}(b) we prove that at late times the structure factor verifies
the dynamical scaling relation  $S(k,t) = R(t)^2 \mathcal F(kR(t))$.

\begin{figure}[t!]
    \centering 
    \includegraphics[width=\columnwidth]{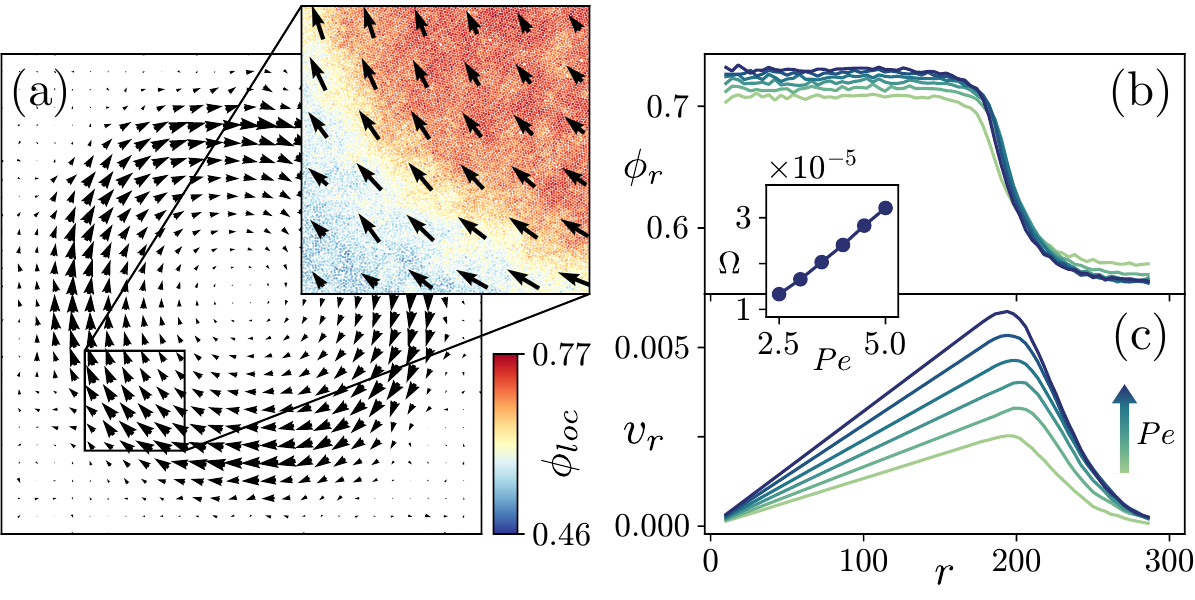}
    \caption{
    (a) Local particle velocity, coarse-grained over a grid of size $20\sigma$ for a dense droplet immersed in the liquid phase.
    The inset shows an enlargement of this field across the interface, on top of the local density.
    Radial profile of (b) the local density and (c) the local velocity, averaged over concentric circular shells, centered in the center of mass of the droplet, for different $Pe$s and clusters of similar radii.
    }
    \label{fig:rotation}
\end{figure}

Finally, as in other chiral systems, we observe in our simulations the presence of edge currents at interfaces [see Fig.~\ref{fig:model}(a)], due to transverse forces acting at dumbbells' encounters, consequence of rotational motion.    
In the particular case of a stationary dense cluster immersed in the liquid phase (\raisebox{-1pt}{\includegraphics[height=0.8\baselineskip]{figures/droplet.pdf}} in Fig.~\ref{fig:phase_diagram}), the cluster is a compact body with frozen relative motion between dumbbells. Transverse active forces sum up to a non zero global torque with respect to the cluster center of mass, resulting in a rigid rotation of the whole cluster [see Fig.~\ref{fig:rotation}(a)]. 
This emerges from the behavior of the circular average of local density and velocity shown in Fig.~\ref{fig:rotation}(b,c).
The cluster rotational velocity $\Omega$ increases linearly with $Pe$, as measured from the slope of the tangential velocity in the interior of the cluster and shown in the inset of Fig.~\ref{fig:rotation}(b).
A theoretical argument for this linear dependence, based on the balance between active and viscous forces, is provided in Appendix D.

In summary, in a minimal system of inertial repulsive dumbbells we have uncovered an emergent macroscopic phase separation under the action of chiral driving and identified its microscopic origin.
Local torque injection raises the translational kinetic energy non-uniformly across density, and the resulting balance between a steric contribution $\alpha(\phi)$ and an inertial contribution $\beta(\phi)$ to the pressure slope generates a genuine region of negative compressibility. 
This mechanism goes quantitatively beyond the instability for inertial spinning disks, which is arrested into finite bubbles rather than driving full separation.
We interpret this contrast as evidence that the instability itself, due to density-dependent kinetic energy storage, is a generic feature of inertial chiral active matter, while the resulting phase morphology is controlled by shape-dependent interfacial properties not captured by the bulk mechanical instability alone. 
Our results also qualify a signature commonly regarded as intrinsic to chiral active systems: edge currents are present at the disordered, liquid-side interfaces of coexisting domains, yet dense clusters instead rotate as rigid bodies, with no internal edge flow. 
Therefore, the presence of edge currents depends on  whether the chiral medium is organized as a fluid or frozen into an ordered structure.

\begin{acknowledgments}
I.P. acknowledges financial support from DURSI under Project No. 2021SGR-673, from MCIU/AEI/FEDER under Grant No. PID2024-156516NB-100, and from the Generalitat de Catalunya through the ICREA Acad\`emia program. 
A.S. acknowledges financial support under the European Union-NextGeneration EU, Mission 4 Component 1, through the Projects of National Relevance Program (PRIN 2022), Grant No. 2022HNW5YL—MOCA.
\end{acknowledgments}

\bibliography{refs}

%\clearpage

\appendix

\onecolumngrid
\vspace{\columnsep}
\begin{center}
\rule[3pt]{0.4\textwidth}{0.4pt}
\textbf{\large{ \ End Matter \ }}
\rule[3pt]{0.4\textwidth}{0.4pt}
\end{center}
\vspace{\columnsep}
\twocolumngrid

\setcounter{equation}{0}
\setcounter{table}{0}
\renewcommand{\theequation}{A\arabic{equation}}

\textit{Appendix A. Dimensionless equations of motion}---
From Eq.~\eqref{eq:langevin}, we derive the dynamics of a single dumbbell's elongation $\bm{d}_i=\mathbf{r}_i-\mathbf{r}_{i+1}$, with $i$, $i+1$ the bead indexes of the dumbbell.
Following the same approach as in~\cite{suma_dumbbellsPRE} and disregarding vibrational degrees of freedom, i.e. assuming that the bead-to-bead distance is fixed $d=\sigma$, we obtain the following equation for the dumbbell's rotation:

\begin{equation}
\label{eq:single_dumbbells_rot}
    I \ddot{\theta} = -\gamma_r \dot{\theta} + \tau_a + \bigl( 2\gamma_r k_BT \bigr)^{1/2} \eta
\end{equation}
where $I$ is the moment of inertia of the dumbbell, $\gamma_r=\gamma \sigma^2/2$ is the rotational friction coefficient, $\tau_a=\sigma F_a$ is the active torque and $\eta$ is a Gaussian white noise of zero mean and unitary amplitude.
Eq.~\eqref{eq:single_dumbbells_rot} describes an inertial rotational diffusion with drift velocity $\omega_a=\tau_a/\gamma_r$ and diffusion constant $D_r=2k_BT/\gamma \sigma^2$.

Similarly, we can derive the equation of motion of the center of mass of a single dumbbell $\mathbf{r}_{cm}=(\mathbf{r}_i+\mathbf{r}_{i+1})/2$.
As pairwise interactions and the active force cancel out, it results a standard diffusion with translational diffusion coefficient $D=k_BT/2\gamma$.

A dimensionless version of Eq.~\eqref{eq:langevin} can be obtained in terms of $(m,\sigma,D_{\rm{r}}^{-1})$ units of mass, lengths and time, respectively.
It reads

\begin{equation}
\label{eq:dless_langevin}
    Pr^{-1} \ \ddot{\mathbf{r}}_i = -\dot{\mathbf{r}}_i - \Gamma \  \boldsymbol{\nabla}_i U + Pe \ \hat{\mathbf{u}}_{\perp,i} + \boldsymbol{\tilde{\xi}}_i \ \rm{,}
\end{equation}
where we have defined a Prandtl number $Pr=\bigl( \frac{\gamma}{m D_r} \bigr)=\bigl( \frac{\gamma \sigma^2}{4m D} \bigr)$, a dimensionless softness $\Gamma=\bigl( \frac{\varepsilon}{2k_BT} \bigr)$ and a rotational P\'eclet number $Pe=\bigl( \frac{\sigma F_a}{2k_BT} \bigr)$.

\setcounter{equation}{0}
\setcounter{table}{0}
\renewcommand{\theequation}{B\arabic{equation}}

\textit{Appendix B. Local density and the binodal}---
The binodal associated to the phase separation at small $Pr$, reported with black triangles in the phase diagram of Fig.~\ref{fig:phase_diagram}, has been identified by a direct measure of the coexistence densities from the probability density functions of the local density $\phi_{loc}$.
We compute $\phi_{loc}$ from the beads' positions as the locally averaged surface fraction over a grid of linear size $20 \sigma$.
They are reported in Fig.~\ref{fig:binodal} for different values of $Pe$ that appear in the phase diagram.
At $Pe=0.5$, for all values of density, we measure a Gaussian distribution peaked at $\phi$.
For $Pe\geq1$, we observe a scenario compatible with the presence of a binodal region within the density range $[\phi_l(Pe):\phi_h(Pe)]$.
For $\phi$ within this range, the system shows a bimodal density distribution with the location of the two peaks corresponding to $\phi_l(Pe)$ and $\phi_h(Pe)$, independent of $\phi$.   

\begin{figure}[t!]
    \centering
    \includegraphics[width=1.\linewidth]{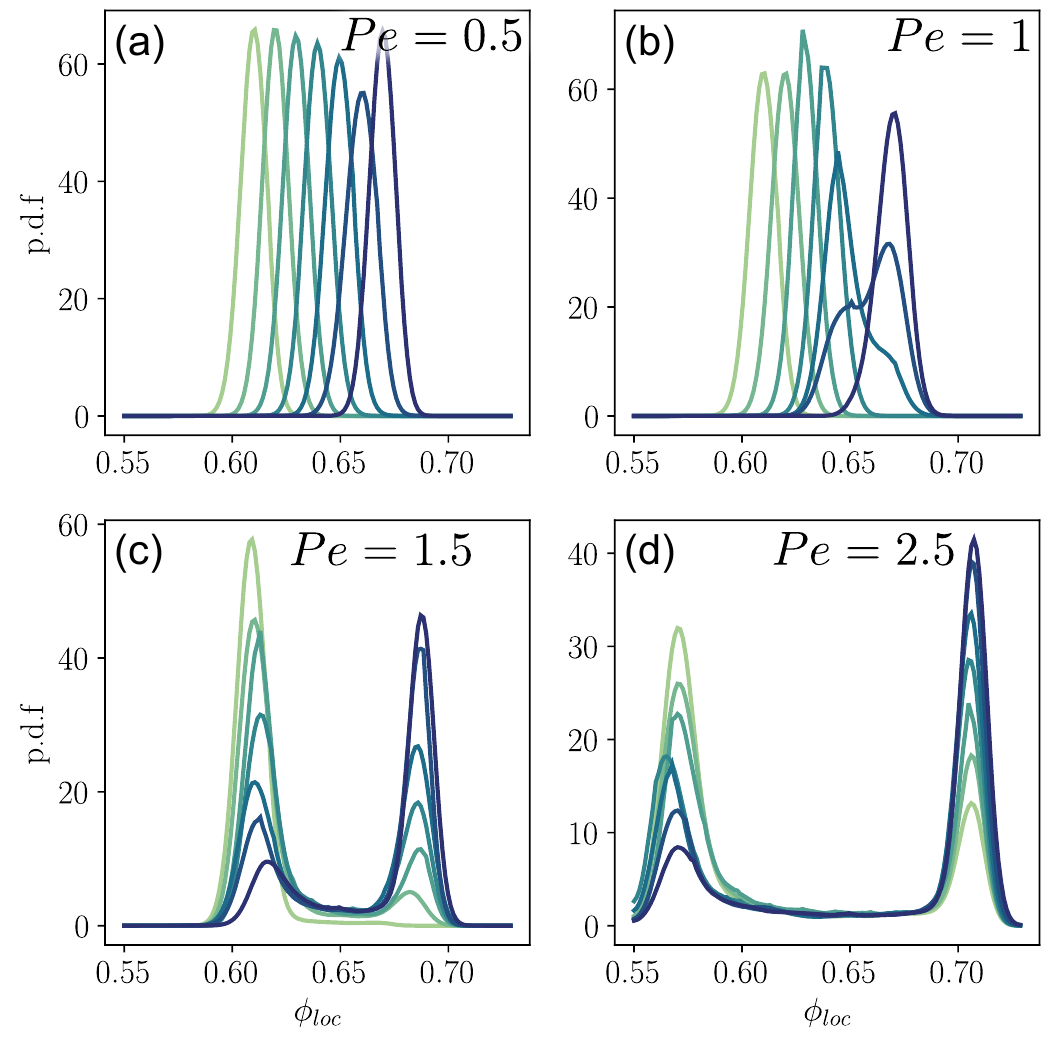}
    \caption{Probability density functions (p.d.f) of the local surface fraction $\phi_{loc}$ for a system of $N=256^2$ bead at different $Pe$s.
    Distributions are computed from independent configurations in the steady regime and three independent simulations.
    }
    \label{fig:binodal}
\end{figure}

\setcounter{equation}{0}
\setcounter{table}{0}
\renewcommand{\theequation}{C\arabic{equation}}

\textit{Appendix C. Small inertia regime}---
While we prove in the main text that inertia is crucial in the appearance of the phase separation, we provide here further evidence of this, exploring the phase behavior within a wide range of the Prandtl number $Pr$.
In Fig.~\ref{fig:phase_diagram_PePrSim}, we report the phase diagram of the system at fixed $\phi=0.66$ and varying $Pr$ and $Pe$.
We observe that increasing $Pr$, thus reducing inertia, the phase separation moves to higher $Pe$, and it is progressively suppressed as we move towards the overdamped limit ($Pr=\infty$).

\begin{figure}[t!!]
    \includegraphics[width=\columnwidth]{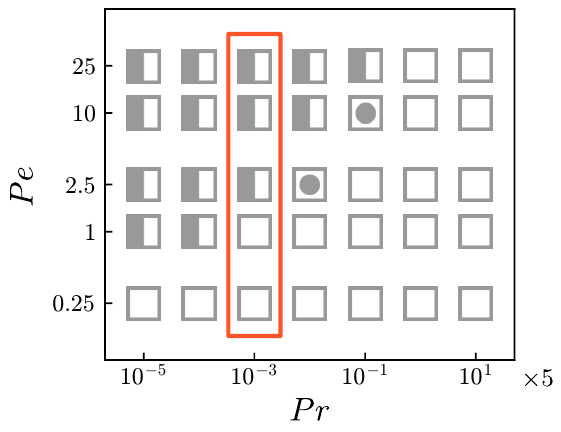}
    \caption{
    $Pr-Pe$ phase diagram at fixed $\phi=0.660$.
    As in Fig.~\ref{fig:phase_diagram}, symbols represent the stationary state of simulations of $N=256^2$ beads, starting from a homogeneous state.
    The red rectangle highlights, as a reference, the set of data points presented in the main text.
    }
    \label{fig:phase_diagram_PePrSim}
\end{figure}

At large $Pr$, dumbbells' translational and rotational inertia becomes negligible, as $\tau_I$ becomes smaller than all the other timescales at play.
As a consequence, the mechanism described in the main text by which the system is able to accumulate part of the energy injected by the active torques is progressively suppressed.
The kinetic energy exchanged by particles at collision is quickly dissipated by friction with the thermal bath and the stationary translational kinetic energy reduces to the bath temperature, independent of the system density [see Fig.~\ref{fig:pressure}(c)].
This prevents the system from developing states of negative compressibility, hence the mechanical instability that drives the phase separation.
We show in Fig.~\ref{fig:Pkin_pr} the kinetic pressure, as defined in the main text, for the system at fixed $Pe=2.5$ and increasing $Pr$, corresponding to the kinetic temperature shown in Fig.~\ref{fig:pressure}(c).
A clear Van der Waals loop is present at $Pr=5 \cdot 10^{-3}$, then it progressively shrinks and disappears for $Pr>0.5$.
This is consistent with the phase diagram of Fig.~\ref{fig:phase_diagram_PePrSim}.

\begin{figure}
    \centering
    \includegraphics[width=\linewidth]{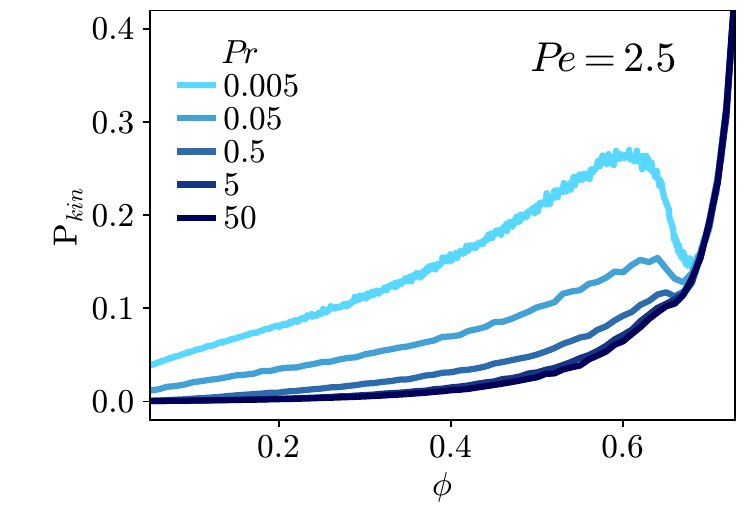}
    \caption{
    Kinetic pressure $P_{kin}$, as defined in the main text, for different values of $Pr$, computed from the kinetic temperature shown in Fig.~\ref{fig:pressure}(c).}
    \label{fig:Pkin_pr}
\end{figure}

\setcounter{equation}{0}
\setcounter{table}{0}
\renewcommand{\theequation}{D\arabic{equation}}

\textit{Appendix D. Rigid rotation of the dense cluster}---
As described in the main text, clusters rotate rigidly about their center of mass with a well-defined angular velocity $\Omega$ [see Fig.~\ref{fig:rotation}]. 
Here we provide a theoretical argument to estimate the dependence of $\Omega$ on the system parameters.
The rigid rotation follows from a balance between the total active torque acting on the cluster and the dissipative torque due to the damping.

The contribution of a single dumbbell is 
\begin{equation}
    (\mathbf{r}_i - \mathbf{r}_{i+1}) \times \mathbf{F}_{i,a} = \sigma F_a  \hat{\mathbf{z}} = \tau_a \hat{\mathbf{z}}
\end{equation}
being $\hat{\mathbf{z}}$ the unit vector perpendicular to the system plane, oriented according to the spinning direction, and we have supposed that bead-to-bead distance is fixed at $|\mathbf{r}_i - \mathbf{r}_{i+1}|=\sigma$, the equilibrium distance of the FENE potential.
The total active torque on a cluster of $N_c$ dumbbells is therefore 
\begin{equation}
    T_a = N_c \tau_a = \rho \pi R^2 \tau_a
\end{equation}
where we have assumed the cluster circular of radius $R$ and $\rho$ is the 2D number density of dumbbells in the cluster. 

As the cluster rotates rigidly with angular velocity $\Omega$, every bead experiences a friction force $-\gamma\Omega r$ depending on its distance $r$ from the center of rotation, thus it contributes with a viscous torque $-\gamma\Omega r^2$. 
Integrating over the cluster gives
\begin{equation}
    T_f = -\gamma \Omega \sum_{i\,\in\,\mathrm{cluster}} r_i^2 \simeq - \gamma\Omega\rho \int_0^R  2\pi r^3\, dr = -\frac{\pi \gamma \Omega \rho}{2} R^4 \ .
\end{equation}

Equating the two opposite torques to have a uniform rotation, we obtain an angular velocity
\begin{equation}
    \label{eq:Omega_vs_Pe}
    \Omega = \frac {2 \tau_a}{\gamma R^2} = \frac{4k_BT}{\gamma R^2} Pe \ \mbox{.}
\end{equation}
This explains the linear dependence on $\Omega \sim Pe$ measured in simulations and reported in Fig.~\ref{fig:rotation}.

\onecolumngrid

%%%%%%%%%% Merge with supplemental materials %%%%%%%%%%
\clearpage
\widetext
\begin{center}
\textbf{\large Supplementary Material for \\
``Phase Separation in a system of Brownian inertial rotors''}

\date{\today}
\end{center}
%%%%%%%%%% Merge with supplemental materials %%%%%%%%%%

\setcounter{equation}{0}
\setcounter{figure}{0}
\setcounter{table}{0}
\setcounter{page}{1}
\makeatletter
\renewcommand{\thefigure}{S\arabic{figure}}

\section{Ordering phase transition}
We provide here further analysis of the phase ordering from the liquid to the solid phase at equilibrium ($Pe=0$) and for $Pe \lesssim 0$.
At $Pe=0$, in the absence of active torques, the system is expected to undergo a Kosterlitz-Thouless-Halperin-Nelson-Young (KTHNY) type ordering scenario, characterized by a phase transition from a isotropic liquid to an hexatic phase, followed by an hexatic-solid phase transition, upon increasing density at constant temperature.
This has been established in~\cite{cugliandolo_dumbbells_2017} for a system of hard rigid dumbbells that features a first-order liquid-hexatic transition and a continuous hexatic-solid one.
Compared to the system used in this work, the dumbbells in~\cite{cugliandolo_dumbbells_2017} interact through a harder repulsive Mie potential $U_{Mie}(r) \sim (\sigma/r)^{64}-(\sigma/r)^{32}$, which allows a smaller overlap between beads.
A harder repulsive core is known to favor a first-order liquid hexatic transition, as shown in~\cite{krauth_continuous_15} for repulsive disks.
In line with this results, we do not observe evidence of first-order transition in our system, thus we conclude that both liquid-hexatic and hexatic-solid are continuous transitions for WCA-interacting dumbbells at equilibrium.

\subsection{Liquid-hexatic transition}

The transition from liquid to hexatic phase is identified by the divergence of the hexatic correlation length and the appearance of quasi-long-range (QLR) hexatic order.
The hexatic correlation function is defined as 
\begin{equation}
    g_6(r) \;=\; \big\langle\,\psi_{6,i}\,\psi_{6,j}^{*}\,
    \big\rangle\big|_{|\mathbf{r}_i-\mathbf{r}_j|=r}
\end{equation}
where $\psi_{6,i}$ is the local hexatic parameter, defined in the main text.
While $g_6$ is exponentially decaying in the isotropic liquid, it becomes as power law $g_6(r) \sim r^{-\eta}$ in the hexatic phase.
The critical value of the exponent $\eta$ predicted by the KTHNY theory is $\eta_c=1/4$.

\begin{figure}[h]
    \centering
    \includegraphics[width=1.\linewidth]{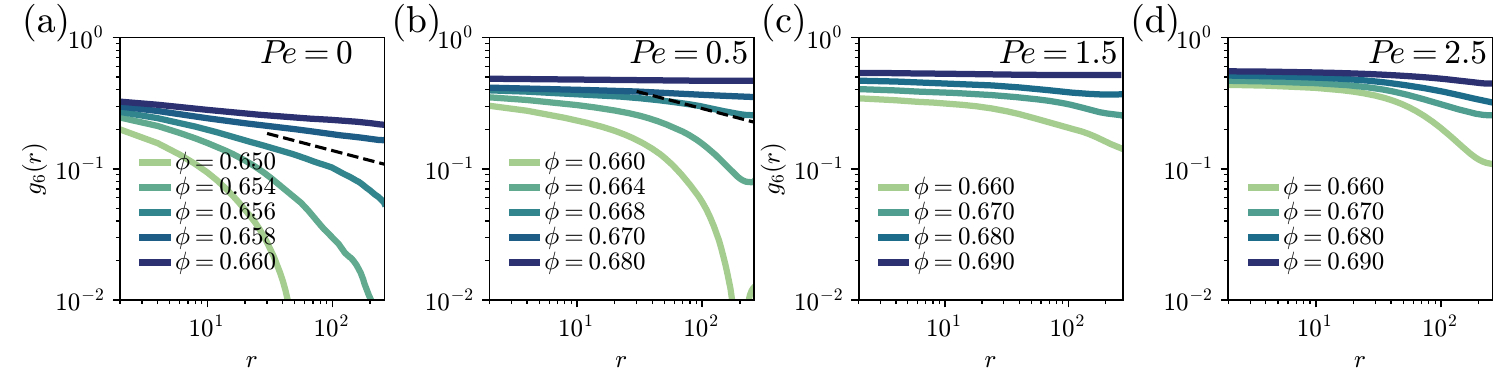}
    \caption{
    Hexatic correlation function $g_6(r)$ for a system of $N_d=512^2/2$ dumbbells at (a) $Pe = 0$, (b) $Pe=0.5$, (c) $Pe=1.5$ and (d) $Pe = 2.5$ for the values of global density $\phi$ indicated in the legends. 
    The dashed line indicates the critical KTHNY power law $r^{-1/4}$, shown as a reference for the liquid-hexatic transition. 
    }
    \label{fig:suppl_hexatic_correlations}
\end{figure}

Fig.~\ref{fig:suppl_hexatic_correlations} reports $g_6(r)$ for a system of $N=512^2$ beads. 
Fig.~\ref{fig:suppl_hexatic_correlations}(a) refers to the passive case $Pe=0$, which shows the presence of a critical point at $\phi_h=0.656$, where the decay of $g_6(r)$ changes for exponential to algebraic.
The slope of the curves close to the transition point is compatible with the KTHNY critical exponent $1/4$.
We observe a similar behavior at $Pe=0.5$ [Fig.~\ref{fig:suppl_hexatic_correlations}(b)], with the critical point moving towards higher density while increasing $Pe$.
In particular, we find $\phi_h=0.668$ at $Pe=0.5$, with an unchanged critical exponent.
The critical points so identified correspond to the ones reported in the phase diagram of Fig.~2 in the main text.

For $Pe>1$ [Fig.~\ref{fig:suppl_hexatic_correlations}(c),(d)] the liquid-hexatic transition competes with the rotation-induces phase separation, with a binodal region stretching over the interval $\phi \in [0.610:0.685]$ for $Pe=1.5$ and $\phi \in [0.565:0.705]$ for $Pe=2.5$.
Consequently, within the binodal region $g_6$ shows a mixed behavior corresponding to the coexistence between the isotropic liquid and a dense hexatic.
This reflects into a slow algebraic decay at small $r$ and a fast exponential decay at large $r$.
The crossover value that separates these two regimes moves to larger $r$ as the system density increases across the coexistence region.

\subsection{Hexatic-solid transition}

Translational order is probed by the correlation function 
\begin{equation}
    C_{\bm{q}_0} = \langle e^{i \bm{q}_0 (\bm{r}_i - \bm{r}_j)} \rangle_{r=|\bm r_i - \bm r_j|} 
\end{equation}
where $\bm q_0$ is the wavevector locating the highest peak of the static structure factor $S(\mathbf{q},t)=N^{-1}\sum_{i,j}^{N}e^{i\mathbf{q}\cdot(\mathbf{r}_i(t)-\mathbf{r}_j(t))}$.
An exponential decay of $C_{\bm{q}_0}$ identifies a phase with short-range translational order (liquid or hexatic), whereas an algebraic decay identifies a solid phase with QLR translational order.

\begin{figure}[h]
    \centering
    \includegraphics[width=1.\linewidth]{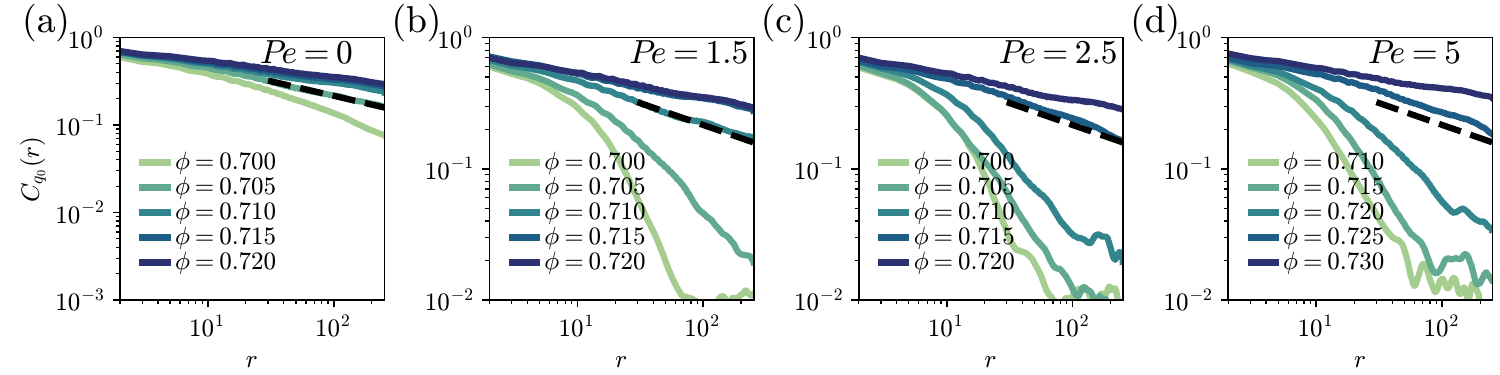}
    \caption{
    Translational correlation function $C_{\bm q_0}$ for a system of $N_d=512^2/2$ dumbbells at (a) $Pe = 0$, (b) $Pe=1.5$, (c) $Pe=2.5$, (d) $Pe=5.0$. 
    The dashed line is the power law $r^{-1/3}$, representing the critical KTHNY power law for the hexatic-solid transition.
    } 
    \label{fig:suppl_transl_correlations}
\end{figure}

Fig.~\ref{fig:suppl_transl_correlations}(a-d) show $C_{\bm{q_0}}$ for $Pe=0,\, 1.5,\, 2.5,\, 5.0$, respectively, spanning densities in the range $\phi \in [0.700:0.730]$. 
For $Pe \leq 2.5$, we observe a continuous hexatic-solid transition, corresponding to a transition from a exponentially to algebraic decays of $C_{\bm{q_0}}$, with a critical exponent equal to the one predicted by the KTHNY theory.
The critical densities are $\phi_s=0.705$ for $Pe=0$, $\phi_s=0.710$ for $Pe=1.5$ and $\phi_s=0.715$ for $Pe=2.5$, reported in the phase diagram shown in Fig.~2 of the main text.

For $Pe=5.0$, as shown in Fig~\ref{fig:suppl_transl_correlations}(d), we observe that for all densities to the right of the binodal---corresponding to the interval $[0.55:0.73]$---the translational correlation function decays algebraically.
For this reason, we conclude that for $Pe \gtrsim 5$ the coexistence is between the isotropic liquid and the solid.

\subsection{Local hexatic order parameter}

To complement the analysis of the phase diagram shown in Fig.~2 of the main text, in Fig.~\ref{suppl: hexproj 0} we report a sequence of snapshots of the projection of the hexatic order parameter $\psi_{6,i}$ along the direction of the global mean for $Pe=0$, $Pe=0.5$ and $Pe=1.5$ and $Pe=5$, at increasing values of the global density $\phi$, spanning the range in which the orientational order develops. 
Orange indicates regions of strong sixfold orientational order, aligned with the principal direction, violet indicates regions of strong sixfold order rotated by $\pi/6$ and intermediate colors indicate either weaker order or different orientations.

\begin{figure}[h]
    \centering
    \includegraphics[width=1.\linewidth]{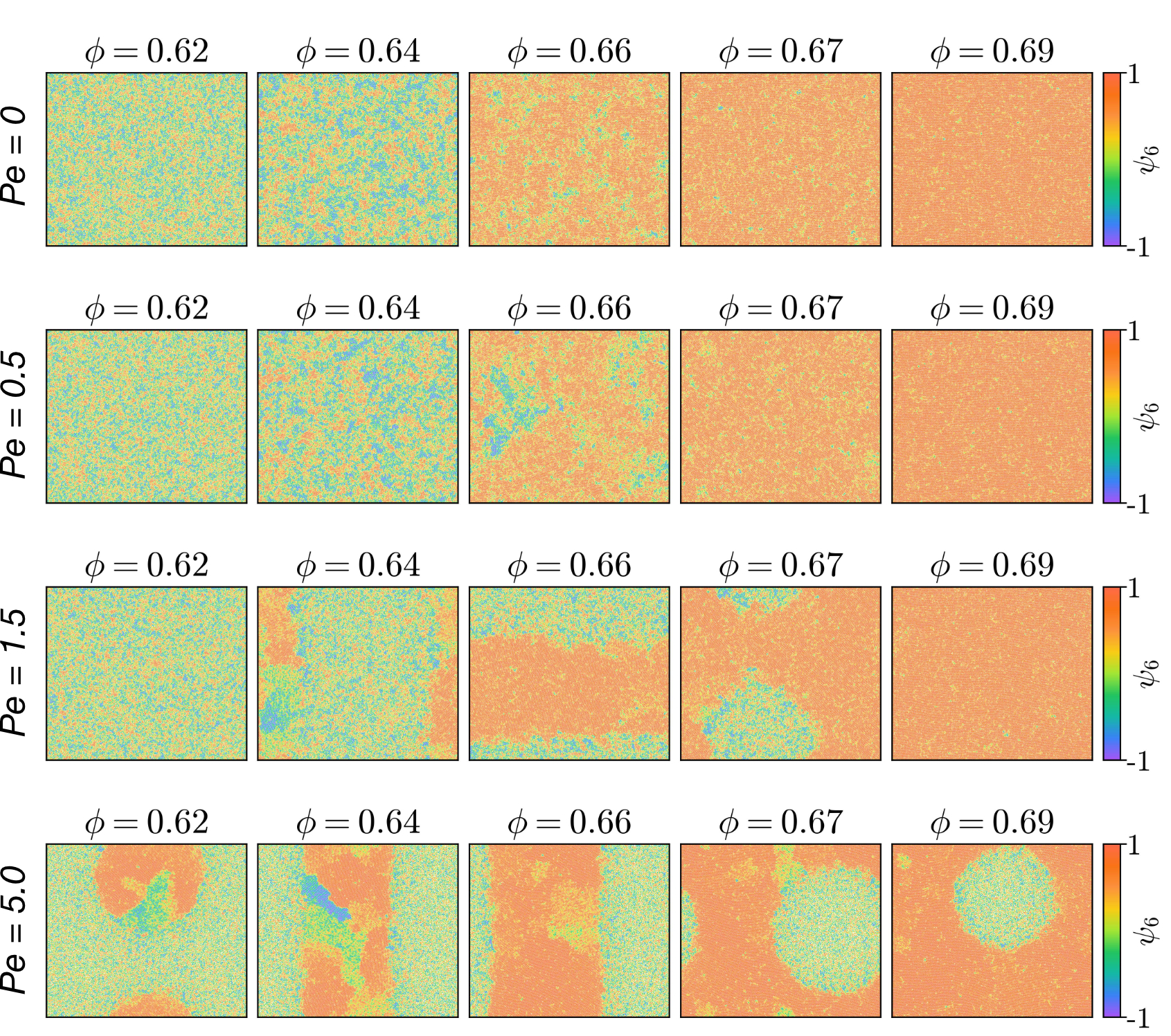}
    \caption{Snapshots of the projection $\psi_6$ of the local hexatic order parameter $\psi_{6,i}$ [see definition in the main text] for $Pe=0, 0.5, 1.5, 5$ at different values of the density.}
    \label{suppl: hexproj 0}
\end{figure}

\section{Irving-Kirkwood pressure for dumbbells}

In the main text the pressure is computed through the Irving--Kirkwood formula
\begin{equation}
\mathrm{P} = \frac{1}{V}\sum_{i=1}^N
\left[ m v_i^2 + \bm{r}_i\cdot\bm{F}_i \right] ,
\label{eq:atomic_pressure}
\end{equation}
which sums over all $N=2N_d$ beads and takes into account all the forces acting on the single beads. 
This \emph{atomic} formulation naturally decomposes into three contributions, $\mathrm{P} = \mathrm{P}_{kin} + \mathrm{P}_{WCA} + \mathrm{P}_{bond}$, where $P_{\mathrm kin}=(1/V)\sum_i m v_i^2$ is the kinetic contribution, computed from the velocities of all beads, $P_{\mathrm WCA}$ is the virial of the repulsive interaction and $P_{\mathrm bond}$ is the virial of the FENE bond.

Equivalently, the pressure can be obtained from a sum running over the $N_d$ dumbbells,
\begin{equation}
    \mathrm{P}^{(m)} = \frac{1}{V}\sum_{a=1}^{N_d}
    \left[M\mathbf{V}_a^{\,2}
        + \mathbf{R}_a \cdot \mathbf{F}_a^{\mathrm{ext}}\right],
    \label{eq:molecular_pressure}
\end{equation}
where $M=2m$ is the mass of the dumbbell, $\mathbf{V}_a$, $\mathbf{R}_a$ and $\mathbf{F}_a^{\mathrm{ext}}$ are the velocity, position and the total external force acting on the dumbbells, respectively. 
By construction, the bond virial does not enter Eq. \ref{eq:molecular_pressure}, as bonds are internal to each molecule. 
Therefore, the kinetic term $\mathrm{P}_{kin}^{(m)}=(M/V)\sum_a \mathbf{V}_a^2$ captures only the translational degrees of freedom, whereas the $\mathrm{P}^{(m)}_{WCA} = \mathbf{R}_a \cdot \mathbf{F}_a$ only includes the inter-molecular WCA interactions. 
The two definitions~\eqref{eq:atomic_pressure} and~\eqref{eq:molecular_pressure} provide alternative coarse-grainings of the same momentum balance and are expected to coincide on average in a homogeneous fluid.

In Fig. 
~\ref{suppl:pressure_atom_mol} we report $\mathrm{P}$, $\mathrm{P}^{(m)}$ and their individual contributions as reported in Eq.~\eqref{eq:atomic_pressure} and Eq.~\eqref{eq:molecular_pressure}, respecrively, at two values of the density $\phi=0.1$ and $\phi=0.7$ and two values of the activity, $Pe=0$ and $Pe=2.5$. 
In all four panels $\mathrm{P}$ and $\mathrm{P}^{(m)}$ coincide on average while displaying a different composition of the individual components.

At low density in the passive limit (top-left panel) the dominant contributions to $P$ are the kinetic term, set by the thermal bath, and the bond virial, which oscillates around its average. 
The WCA virial is small, because collisions between distinct dumbbells are rare. 
The total $P$ inherits the oscillation of the bond virial.
The molecular pressure $P_{\rm mol}$ by construction does not include the FENE contribution.

Also at $\phi=0.1$ and $Pe=2.5$ the two formulations agree on average, with different contributions.
In the atomic pressure, $\mathrm P_{kin}$ is dominated by the rotational kinetic energy of the beads and is balanced by a contractive bond virial $\mathrm P_{bond}<0$ that supplies the centripetal force; the WCA virial remains small as collisions are rare.
Although interactions between dumbbells are infrequent, as seen from the negligible contribution to the total pressure of the WCA virial $\mathrm{ P}^{(m)}_{WCA}$, they are responsible for converting rotational kinetic energy into translational energy, resulting in an increased value of the kinetic pressure $\mathrm{P}_{kin}^{(m)}$.

At $\phi=0.7$, the repulsive potentials dominate in both formulations and the role of the activity becomes marginal. 
In the atomic decomposition $\mathrm P\simeq \mathrm P_{\rm WCA}+\mathrm P_{\rm bond}$, with $\mathrm P_{\rm bond}$ giving a small but non negligible correction. 
In the molecular decomposition the bond virial is absent and the pressure is mostly set by $\mathrm P_{\rm WCA}^{(m)}$ alone.

\begin{figure}
    \centering
    \includegraphics[width=0.8\linewidth]{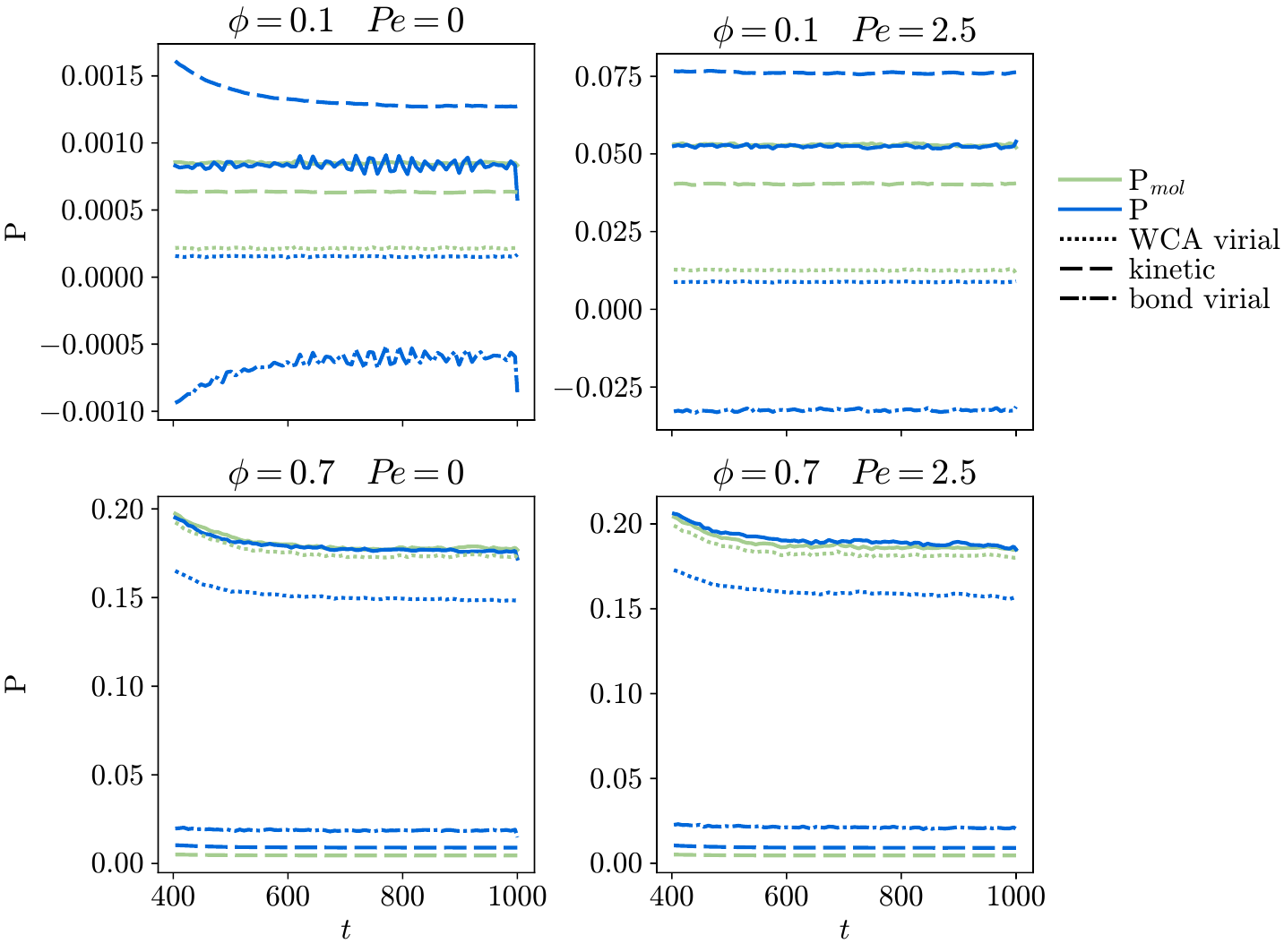}
    \caption{
    Time series of the atomic pressure $P$, of the molecular pressure $P^{(m)}$ and their respective individual contributions. The four panels correspond to two values of the density ($\phi=0.1$, top, and $\phi=0.7$, bottom) and two values of the activity ($Pe=0$, left, and $Pe=2.5$, right).}  \label{suppl:pressure_atom_mol}
\end{figure}

%\bibliography{refs}

\end{document}